\documentclass[onecolumn,11pt,preprintnumbers,floatfix,showpacs,prl,superscriptaddress,reprint]{revtex4-1}
\usepackage{graphicx}% Include figure files
\usepackage{dcolumn} % Align table columns on decimal point
\usepackage{bm}
\usepackage{epsfig}
\usepackage{longtable}
\begin{document} 

\title{Charge ordering at the interface in (LaMnO$_3$)$_{2n}$/(SrMnO$_3$)$_n$ superlattices as the origin of their insulating state}

\author{V\'ictor Pardo}
\affiliation{Departamento de F\'{i}sica Aplicada,
  Universidade de Santiago de Compostela, E-15782 Campus Sur s/n,
  Santiago de Compostela, Spain}
\affiliation{Instituto de Investigaci\'{o}ns Tecnol\'{o}xicas,
  Universidade de Santiago de Compostela, E-15782 Campus Sur s/n,
  Santiago de Compostela, Spain}  
  \email{victor.pardo@usc.es}
\author{Ant\'ia.~S. Botana}
\affiliation{Departamento de F\'{i}sica Aplicada,
  Universidade de Santiago de Compostela, E-15782 Campus Sur s/n,
 Santiago de Compostela, Spain}
 \affiliation{Instituto de Investigaci\'{o}ns Tecnol\'{o}xicas,
  Universidade de Santiago de Compostela, E-15782 Campus Sur s/n,
  Santiago de Compostela, Spain}  
\author{Daniel Baldomir}
\affiliation{Departamento de F\'{i}sica Aplicada,
  Universidade de Santiago de Compostela, E-15782 Campus Sur s/n,
  Santiago de Compostela, Spain}
\affiliation{Instituto de Investigaci\'{o}ns Tecnol\'{o}xicas,
  Universidade de Santiago de Compostela, E-15782 Campus Sur s/n,
  Santiago de Compostela, Spain}

\date{\today}

\begin{abstract}

We have performed ab initio calculations within the LDA+U method in the multilayered system (LaMnO$_3$)$_{2n}$ / (SrMnO$_3$)$_n$. Our results suggest a charge-ordered state that alternates Mn$^{3+}$ and Mn$^{4+}$ cations in a checkerboard in-plane pattern is developed at the interfacial layer, leading to a gap opening.  Such an interfacial charge-ordered situation would be the energetically favored reconstruction between LaMnO$_3$ and SrMnO$_3$. This helps understanding the insulating behavior observed experimentally in these multilayers at intermediate values of $n$, whose origin is known to be due to some interfacial mechanism.

\end{abstract}

\maketitle

%\section{BACKGROUND}

Recently focus has been drawn on the intriguing metal-insulator transition that occurs in (LaMnO$_3$)$_{2n}$ / (SrMnO$_3$)$_n$ superlattices as a function of the parameter $n$ controlling the thickness of each sublayer.\cite{lmo_smo_mit_anand,lmo_smo_mit_adamo_apl,lmo_smo_mit_dagotto,lmo_smo_nmat} The interest in this particular 2$n$/$n$ multilayer arrangement would be to study the progress towards the La$_{2/3}$Sr$_{1/3}$MnO$_3$ (LSMO) solid solution that shows a large negative magnetoresistance and other interesting physical properties.\cite{manganites_cmr}  Intuitively one could think that for short periods $n$, the system would behave in a similar way to the solid solution: with the standard interfacial roughness on the order of one unit cell, the system would be similar to a solid solution. On the opposite limit, for large $n$, one could think of two distinct insulating blocks formed by each of the constituents: SrMnO$_3$ (SMO) and LaMnO$_3$ (LMO). The intermediate-$n$ regime where the transition from localized to itinerant behavior should occur is not that clear but many interesting physical phenomena can arise and the physics at the interface is expected to play a big role.

SMO\cite{smo_bulk} is a cubic perovskite formed by Mn$^{4+}$:d$^3$ cations where the full t$_{2g}$ shell leads to an isotropic G-type antiferromagnetic (AF) insulating state. LMO\cite{lmo_bulk} is a distorted A-type antiferromagnet with a peculiar type of orbital ordering\cite{lmo_bulk_oo} of the only occupied e$_{g}$ state of the Mn$^{3+}$:d$^4$ cations, the system showing a prononunced Jahn-Teller distortion. If the interface between these LMO and SMO blocks in their multilayers were insulating, the whole system should then be insulating. However, it is known that when two perovskite oxides, one La-based and the other Sr-based are put together along the (001) direction, like the case studied here, the polar nature of the interface is prone to show metallicity or other types of electronic reconstruction.\cite{lao_sto_hwang,lao_sto_pickett} 
 
At small thicknesses, the interface between SMO and LMO (or between the solid solution and STO) is polar and hence important reconstructions may appear. Also, the interfacial Mn atoms have both La$^{3+}$ and Sr$^{2+}$ nearest neighbors, their average valence is hence non-integer Mn$^{3.5+}$ and that leads to partially filled bands with the overall system being metallic. 
However, experimental evidences in the (LMO)$_{2n}$/(SMO)$_n$ superlattices show that there is a critical thickness for metallicity to stop showing up, at about $n$= 3. 
Hence, the large-$n$ limit is insulating and the small-$n$ limit is metallic. The reason for the transition has been ascribed to be some kind of interfacial disorder leading to carrier localization\cite{lmo_smo_mit_anand} but also experiments have shown the key role played by the interface in the transport properties,\cite{lmo_smo_mit_adamo_apl,lmo_smo_adamo_interfaces} where the different range of effective dopings and charge transfer through the interface can be dominant.\cite{lmo_smo_magnetism_adamo}

This type of transition is not unique to the 2$n$/$n$ multilayers, it also occurs for the nominal La$_{0.5}$Sr$_{0.5}$MnO$_3$ concentration of the equivalent solid solution (the case of (LMO)$_n$/(SMO)$_n$ multilayers), even though the critical $n$ can vary slightly in that case, $n$ being larger.\cite{lmo_smo_05} Also, it has been shown recently that if a solid solution La$_{2/3}$Sr$_{1/3}$MnO$_3$ film is grown on top of SrTiO$_3$(STO), a similar effect happens, namely at small thicknesses the system becomes insulating,\cite{benjamin_prl} with very similar resistivity vs. temperature curves as a function of film thickness to those reported in the multilayered system as a function of $n$. But the multilayered structure is not needed in that case, it all can be explained as an effect of dimensionality reduction and the corresponding structural changes and octahedral distortions the manganite undergoes when grown epitaxially in very thin films on top of STO due to the strain induced by the substrate.

Motivated by the possible appearance of all these effects, it is interesting to address the point of what are the interfacial reconstructions that occur, and that is the goal of the present Letter. 
%At small $n$, the interfaces dominate the overall conductivity. The question is what happens at intermediate $n$ values where the transition occurs.
Ab initio calculations have been performed in the past in this multilayered system.\cite{satpathy_prb_08,satpathy_prb_09,satpathy_prb_10} For a small sublayer thickness, the interfacial Mn atoms have an intermediate average valence and the system as a whole would be metallic if no additional reconstructions take place. As the $n$ value is increased, those calculations\cite{satpathy_prb_09} show a dip in the DOS appearing  but never a gap opening. None of the mechanisms put forward from a band structure point of view for the physics at the interface can get completely full bands for the interfacial layers.

Here, we propose a different picture for the physics at the interface to explain why at intermediate film thicknesses, the system can be insulating. Our calculations suggest that the interface can develop a charge-ordered layer between insulating bulk-like SMO and LMO and that is the preferred interfacial reconstruction, leading to an insulating interfacial layer. The average mixed valence of interfacial Mn atoms (typically Mn$^{3.5+}$) can order accross the interface to yield a charge-ordered layer Mn$^{3+}$/Mn$^{4+}$ that produces a gap opening and explains the experimental observations. This type of ordering occurs at half-filling in bulk La$_{1-x}$Ca$_x$MnO$_3$\cite{co_vpardo, co_volja} compound and the whole phase diagram of LSMO shows charge-ordered phases at different compositions.\cite{co_lsmo_experiments} Charge ordering (CO) is at the heart of various prominent metal-insulator transitions in oxides\cite{prl_co} and has been found in various other perovskite-based oxide interfaces systems, such as LaAlO$_3$/SrTiO$_3$\cite{lao_sto_pentcheva_co} or LaNiO$_3$/LaAlO$_3$.\cite{lno_lao_pentcheva_co,new_lno_lao} Also, several of these mixed-valent manganites show charge-ordered states in their bulks. However, the strength of the CO state depends on the structural details.\cite{lmo_smo_tco} LSMO being rhombohedral in its bulk state lacks the possibility for Jahn-Teller distortions required for CO to occur, but Ref. \onlinecite{benjamin_prl} shows that when grown in thin films on top of STO, such lattice instabilities are possible due to the symmetry reduction.

%\section{COMPUTATIONAL PROCEDURES}

Our electronic structure calculations were performed within density functional
theory\cite{dft,dft_2} using the all-electron, full potential code {\sc wien2k}\cite{wien} based on the augmented plane wave plus local orbitals (APW+lo) basis set.\cite{sjo}  To deal with strong correlation effects,
we apply the LDA+$U$ scheme \cite{sic} that incorporates an on-site Coulomb repulsion $U$ and Hund's rule coupling strength $J_H$ 
for the Cr $3d$ states. The LDA+$U$ scheme improves over GGA or LDA in the study of systems containing correlated electrons by introducing the on-site Coulomb repulsion $U$ applied to localized electrons. The uncorrelated part of the potential was modelled using the local density approximation (LDA).\cite{lda} We have performed calculations taking various $U$ values for Mn, in the range 4-7 eV. The on-site Hund's exchange parameter $J_H$ was set to 0.75 eV, values in the usual range for Mn 3d electrons. We have performed our calculations (unless stated for comparisons) fixing the in-plane lattice parameter to that of SrTiO$_3$, a typical substrate to grow these interfaces on, which has a reasonable lattice match with SMO. We have fully relaxed volume (out of plane) and internal coordinates of the system using the LDA+$U$ scheme, which allows to relax the Jahn-Teller distortions associated with the d$^4$ electron count in Mn$^{3+}$ cations. All calculations were fully converged with respect to all the parameters used. In particular, we used R$_{mt}$K$_{max}$= 6.0, 
and muffin-tin radii of 2.19 for Sr, 2.31 for La, 1.82 a.u. for Mn and 1.61 a.u. for O.

%\section{RESULTS}

\begin{figure}
\includegraphics[width=\columnwidth,draft=false]{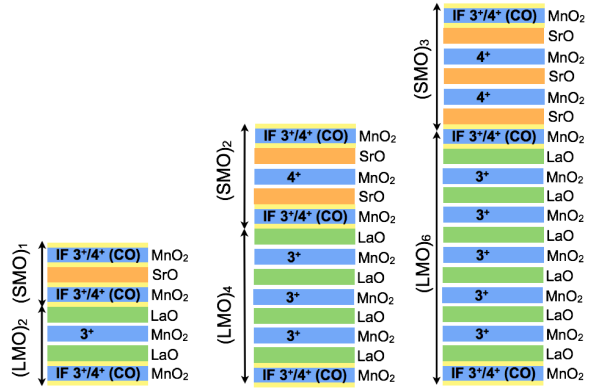}
\caption{(Color online) Schematic representation of the (LMO)$_{2n}$/(SMO)$_n$ multilayered system as $n$ increases from 1 on the left to 2 in the middle and 3 on the right. A charge-ordered Mn$^{3+}$/Mn$^{4+}$ layer appears separating two insulating LMO and SMO sublayers.}\label{full}
\end{figure}

Figure \ref{full} shows a schematic representation of the typical structure of the (LMO)$_{2n}$/(SMO)$_n$ superlattices and the main interfacial reconstruction that our calculations show. The interfacial Mn atoms are surrounded by both La and Sr, hence their average valence is non-integer, Mn$^{3.5+}$: d$^{3.5}$, thus it would be metallic had charge order (or other possible reconstruction) not taken place. The $n$= 1 system does not even develop an SMO-only layer, which is one-atomic-layer thick for $n$= 2 (probably comparable to the standard ionic interdiffusion in this kind of interfaces) and becomes two-layer thick for $n$= 3.

\begin{figure}
\includegraphics[width=\columnwidth,draft=false]{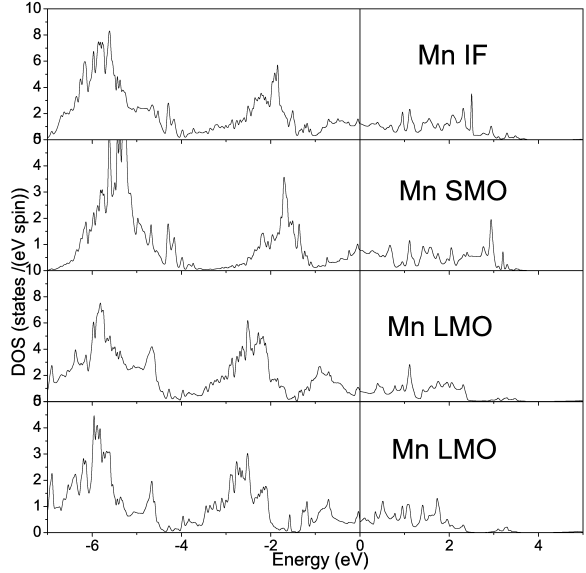}
\caption{(Color online) Layer-resolved DOS of the Mn atoms for the $n$= 2 system without charge order. The solution is metallic. Only the majority spin channel is shown, the minority one being fully unoccupied in every case.}\label{metallic}
\end{figure}

It was seen in the past that a gap opening situation cannot be obtained via ab initio calculations. Ref. \onlinecite{satpathy_prb_09} has shown that as $n$ increases a dip at the Fermi energy can be seen but no gap opens up, this only happens at large $n$ for the non-interfacial layers, the system being metallic overall.  Our calculations confirm these results: if a charge ordered layer is not imposed at the interface the system is metallic (even trying various types of magnetic orderings, which do not seem crucial - see below - for that sake). This can be seen in Fig. \ref{metallic} where the layer-resolved DOS  for a (LMO)$_4$/(SMO)$_2$ multilayer is shown. A solution without CO leads to a a metallic behavior.

\begin{figure}
\includegraphics[width=\columnwidth,draft=false]{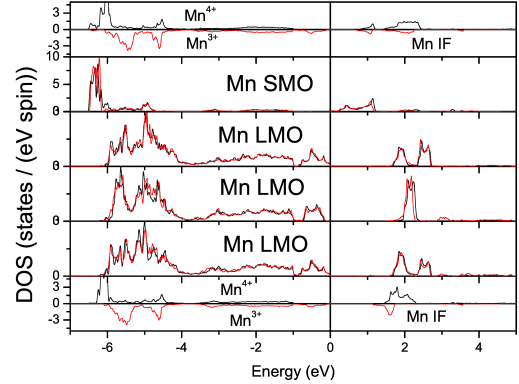}
\caption{(Color online) Layer-resolved density of states of the charge-ordered multilayer system ($n$= 2). A gap opens at the Fermi level once charge order is set. For each layer, two Mn atoms with opposite spins are shown as positive or negative.}\label{dos_2}
\end{figure}

\begin{figure}
\includegraphics[width=\columnwidth,draft=false]{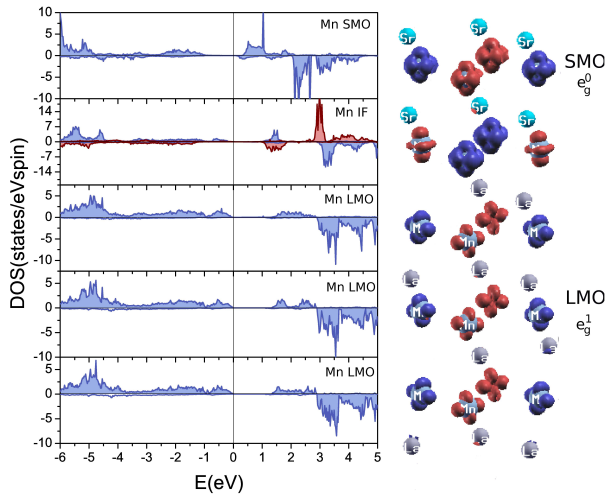}
\caption{(Color online) Layer-resolved density of states (left) of the charge-ordered multilayer system ($n$= 3). Positive (negative) values indicate majority (minority) spin. A gap opens at the Fermi level once charge order is set. On the right, a spin-density plot showing the unoccupied part of the Mn e$_g$ spectrum. A single unoccupied e$_g$ band corresponds to Mn$^{3+}$: d$^4$ cations, whereas the combination of two e$_g$ orbitals indicates a Mn$^{4+}$:d$^3$ ion. The interfacial layer shows Mn$^{3+}$/Mn$^{4+}$ alternation. Different colors represent different spin channels.}\label{dos}
\end{figure}

The scenario is different if an interfacial charge-ordered layer can be formed, separating well-defined SMO and LMO blocks, both insulating. Such a situation will lead to a gap opening in the whole system. In Figs. \ref{dos_2} and \ref{dos} we show the layer-resolved DOS plots for the $n$= 2 and $n$= 3 cases, respectively, when CO is imposed at the interface. That is the ground state on an LDA+$U$ scheme for the reasonable values of $U$ from 4 to 7 eV that we have tried. 
When the internal coordinates are fully relaxed using LDA+$U$, a charge-ordered solution is lower in energy and a gap opens at the Fermi level.  This ground state can be stabilized both in an orthorhombic unit cell and also when the in-plane lattice parameters are fixed to those of SrTiO$_3$ (independently on the off-plane volume optimization), as long as the Jahn-Teller distortion of the Mn$^{3+}$: d$^4$ cations is permitted. This is a crucial point, symmetry needs to be low enough to allow for the distortions that give rise to the peculiar orbital ordering observed experimentally in LaMnO$_3$ in order to obtain an insulating solution. The Mn-Mn out-of-plane distances are, once relaxed: 3.84 \AA\ for the LMO layer, 3.69 \AA\ for the SMO layer and 3.77 \AA\ for the interfacial charge ordered layer. These distances (calculated with LDA+U for U= 5 eV) relax within one unit cell inside the LMO or SMO layers.

In order to understand and pictorially show the interfacial CO, we plot the unoccupied majority-spin density (see the right panel of Fig. \ref{dos}), i.e. the part of the majority-spin e$_g$ spectrum that is unoccupied for each Mn cation (for Mn$^{3+}$:t$_{2g}$$^3$e$_g$$^1$ only one missing e$_g$ electron will appear in the plot and for Mn$^{4+}$:t$_{2g}$$^3$e$_g$$^0$ two missing e$_g$ electrons will appear). One can distinguish three zones: i) an LMO-like zone where $U$ together with the local environment typical of LMO allows for orbital ordering to occur and the splitting of the e$_g$ doublet leads to a gap opening; this LMO layer (formed by Mn$^{3+}$: d$^4$ cations) appears with just one electron in the unoccupied part of the spectrum,  ii) an SMO-like region (formed by Mn$^{4+}$: d$^3$ atoms), with a t$_{2g}^3$-e$_g^0$ gap in the DOS and in the spin-density plot showing two electrons in the unoccupied part of the spectrum, and iii) an interfacial layer that is charge ordered and mixes both types of Mn atoms in a checkerboard fashion. The energy difference with a non-CO metallic solution is always larger than 100 meV/Mn (very large in every case analyzed). This interfacial mechanism could explain why starting at $n$= 3, these multilayers stop showing any conductivity coming from the interface.

The situation is similar to the $n$= 2 case. This suggests that as soon as an interface can occur with distinct LMO and SMO layers plus an interfacial layer, charge order is stable at the interface. Locally, Mn with an average valence 3.5+ will develop a checkerboard AF charge-ordered structure. We have also performed calculations for $n$= 1 and also for various (SMO)$_1$/(LMO)$_m$ systems, and in every case the preferred interfacial reconstruction is forming an insulating charge ordered interface. The fact that an insulating $n$= 2 case is not observed experimentally could have to do with the standard (at least) one-unit-cell imperfections in the interface between the two subsystems. That is why the resistivity curve does not yet develop an insulating character. These imperfections would destroy our picture, whereas the thickness increases, an interface separating a well-defined SMO (or LMO) sublayer can be formed, leading to an insulating charge ordered interface, that according to our calculations is always stable in the perfectly ordered system.

Figure \ref{dos} shows also the type of in-plane checkerboard magnetic ordering imposed. However, as mentioned above the kind of magnetic ordering is not crucial for the gap opening at the interface, because the calculation shown in Fig. \ref{metallic} also has the same magnetic structure and the solution is metallic. It is CO that matters, and it becomes largely stable. Magnetism was analyzed in the past\cite{satpathy_prb_09, satpathy_prb_10} with various types of magnetic couplings being studied, none of them leading to a gap opening. We have studied various types of magnetic orderings and the ground state is stabilized by an A-type AF ordering in the LMO sublayers, G-type in the SMO layer (when thick enough to make it possible) and in-plane checkerboard AF ordering in the charge ordered interfacial layer.

To summarize, our work suggests a charge-order-based mechanism for the interfacial reconstruction in a polar interface between two Mn-based oxides. This allows to understand why the (LMO)$_{2n}$/(SMO)$_n$ multilayered system becomes insulating for $n$ $=$ 3 because the nominally mixed valent interface becomes charge ordered and a gap appears around the Fermi level.

%\section{Summary}

%\acknowledgments

The authors thank the Ministerio de Educaci\'{o}n y Ciencia (MEC) and Xunta de Galicia for the financial support through the projects MAT2009-08165 and EM2013/037. V. Pardo and A.~S. Botana thank the Spanish Government for financial support through the Ram\'on y Cajal and FPU Program, respectively. We thank F. Rivadulla for fruitful discussions.

%\bibliography{lmo_smo}

%merlin.mbs aipnum4-1.bst 2010-07-25 4.21a (PWD, AO, DPC) hacked
%Control: key (0)
%Control: author (8) initials jnrlst
%Control: editor formatted (1) identically to author
%Control: production of article title (-1) disabled
%Control: page (0) single
%Control: year (1) truncated
%Control: production of eprint (0) enabled

%

\end{document}